\documentclass[preprint,12pt,authoryear]{elsarticle}
\setcitestyle{numbers} 

\usepackage{amssymb}

\usepackage{algorithmic}
\usepackage{graphicx}
\usepackage{textcomp}
\usepackage{xcolor}
\usepackage{verbatim}
\usepackage{caption}
\usepackage{subcaption}
\usepackage{multirow}
\usepackage{color, colortbl}
\usepackage{lscape}
\usepackage{placeins}
\usepackage{threeparttable}
\usepackage{amsmath}

\journal{Journal of Parallel and Distributed Computing}

\begin{document}

\begin{frontmatter}



\title{HALO 1.0: A Hardware-agnostic Accelerator Orchestration Framework for Enabling Hardware-agnostic Programming with True Performance Portability for Heterogeneous HPC}


\author{Michael Riera, Erfan Bank Tavakoli, Masudul Hassan Quraishi, Fengbo Ren}

\address{School of Computing and Augmented Intelligence, Arizona State University, Tempe, Arizona, USA}

\begin{abstract}
This paper presents HALO 1.0, an open-ended extensible multi-agent software framework that implements a set of proposed hardware-agnostic accelerator orchestration (HALO) principles. HALO implements a novel compute-centric message passing interface (C$^2$MPI) specification for enabling the performance\nobreakdash-portable execution of a hardware-agnostic host application across heterogeneous accelerators. The experiment results of evaluating eight widely used HPC subroutines based on Intel Xeon E5-2620 CPUs, Intel Arria 10 GX FPGAs, and NVIDIA GeForce RTX 2080 Ti GPUs show that HALO 1.0 allows for a unified control flow for host programs to run across all the computing devices with a consistently top performance portability score, which is up to five orders of magnitude higher than the OpenCL-based solution.
\end{abstract}




\end{frontmatter}


\section{Introduction}
High-performance computing (HPC) applications have been becoming increasingly complex in recent years \cite{10.1145/3295500.3356197,10.1145/3295500.3356209,10.1109/SC.2018.00013,10.1109/SC.2018.00015,10.1109/SC.2018.00044, 10.1109/SC.2018.00053}. Predictive simulations with increasingly higher spatial and temporal resolutions and ever-growing degrees of freedom are the critical drivers for achieving scientific break-through \cite{10.1145/3295500.3356202,10.1145/3229710.3229732,10.1145/3295500.3356147,10.1145/3295500.3356155,10.1145/3295500.3356180}. The latest advancements in deep learning paired with the next-generation scientific computing applications will inevitably demand orders of magnitude more compute power for future HPC infrastructure. In the concluding days of Moore's law, general-purpose solutions will no longer be viable for continuing to meet such an exponential growth in HPC performance that is required to keep pace with scientific innovations \cite{10.1109/SC.2004.68,10.1145/1188455.1188684,10.1145/1328554.1328556}. We envision that extreme-scale heterogeneous HPC systems that massively integrate various domain- and application-specific accelerators will be a viable blueprint for providing the necessary performance and energy efficiency to meet the challenges of future sciences. 

\begin{figure}[]
  \centering
  \includegraphics[]{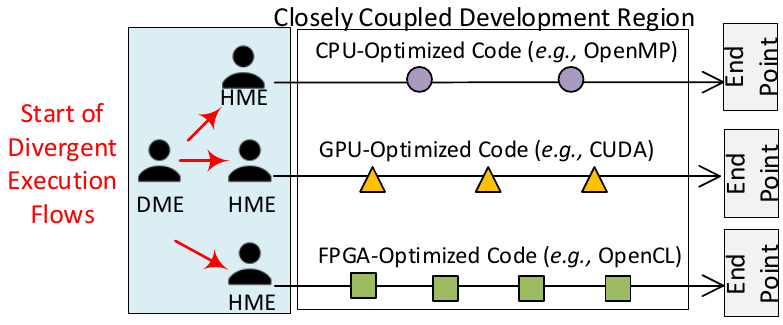}
  \caption{Illustration of divergent execution flows of hardware-optimized application codes in the existing HPC systems. Domain-matter experts (DME) and Hardware-matter experts (HME) must work interdependently in a closely coupled development region. }
  \label{Figure: tightly coupled HPC code}
\end{figure}

However, the path to realizing extreme-scale heterogeneous HPC is tortuous \cite{10.1145/2063348.2063374,10.5555/3018088.3018090, 10.5555/3018088.3018092, 10.5555/3018088.3018094,10.1145/2831425.2833197}. The main obstacle towards the proliferation of heterogeneous accelerators is the lack of a flexible hardware-agnostic programming model that minimizes the dependency of hardware-specific logic throughout the host code while unifying the programming model for scaling up and out the application. As a result, existing HPC applications are by no means extensible with regard to new accelerator hardware. With the lack of clarity in the demarcation between hardware-specific and hardware-agnostic development regions (see Figure \ref{Figure: tightly coupled HPC code}), today's programming models require domain-matter experts (DMEs) and hardware-matter experts (HMEs) to work interdependently to make a significant effort in optimizing hardware-specific codes in order to adopt new accelerator devices in HPC and gain performance benefits \cite{10.1145/3204919.3204937, 10.1145/1188455.1188673, 10.1109/WACCPD.2014.11}. Such a tangled association is a self-imposed bottleneck from our existing programming models that impairs the future of extreme-scale heterogeneous HPC and severely impacts the velocity of scientific discovery.

 We envision that hardware-agnostic programming with high performance portability will be the bedrock for realizing the pervasive adoption of emerging accelerator technologies in future heterogeneous HPC systems. Hardware-agnostic programming is the methodology paired with a programming model that enables DMEs to focus on orchestrating data to and from hardware-specific kernels without being bogged down by the intricacies of target hardware. Data orchestration includes the conditioning (reformatting) and steering of data within the host code. 
We propose to enforce a clear separation between hardware-agnostic host code programming and hardware-specific kernel code optimization, through which we aim to maximize the performance portability of host codes as the first step to minimizing the hardware dependency of HPC applications, as well as maximizing the reusability of kernel codes for reducing unnecessary code refactoring efforts. 
Additionally, we define performance portability in the strictest sense as the ability for the host code to maintain a unified hardware-agnostic control flow, as well as state-of-the-art kernel performance, regardless of platform and/or scale. Additionally, performance portability includes the ability to dynamically handle various accelerators without recompilation of the host code. This is in stark contrast to the conventional definition that allows for multiple control flows and recompilation processes. 
 
 The existing solutions to portability range from programming models (\textit{e.g.}, OpenCL \cite{10.1145/3388333.3388658, 10.1109/ScalA.2014.8, chen2012using}, OpenMP \cite{dagum1998openmp}, Thrust \cite{10.5555/3314872.3314914} , and OpenACC \cite{10.1109/CCGrid.2013.12, 10.1109/WACCPD.2014.10}) to software libraries and frameworks (\textit{e.g.}, MKL \cite{10.1145/306113.306123}, MAGMA \cite{10.1145/2664666.2664667}, HIPSyCL \cite{10.1145/3388333.3388641}, KOKKOS \cite{10.1145/3318170.3318193}, RAJA \cite{10.5555/3314872.3314914} ) \cite{10.1145/3204919.3204939}. However, these solutions are mostly built to provide functional portability rather than performance portability \cite{reinders2021data, https://doi.org/10.1002/cpe.5640, 10.1109/CCGrid.2013.12, 10.1145/2807591.2807621, 10.1145/3110355.3110356}. In addition, all these solutions lack one or more critical components that are necessary for providing host code hardware agnosticism, transparent interoperability, simplicity, and/or extensibility. Additionally, current solutions require multiple code paths to be embedded into the host code for supporting multiple accelerators, creating a highly hardware-dependent application that is hardly manageable beyond two accelerators (see Figure \ref{Figure: tightly coupled HPC code}). Furthermore, current solutions lack the scaling capabilities required at scale, especially exascale.

\begin{figure}
\centering
  \includegraphics[scale=.5]{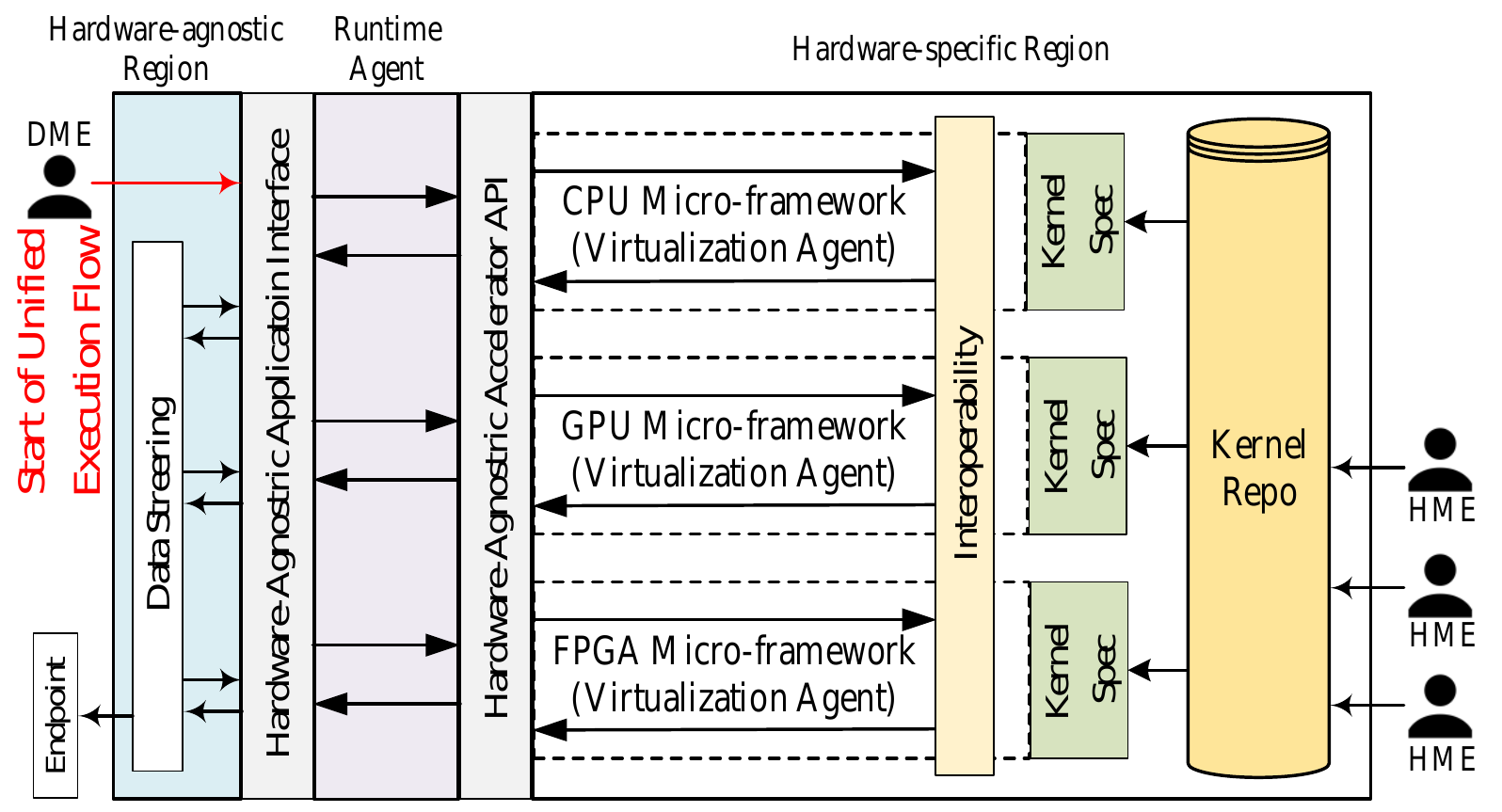}
  \caption{Illustration of a unified execution flow of hardware-agnostic application codes enabled by HALO 1.0. DMEs and HMEs can work independently in completely decoupled development regions.}
  \label{Figure: HALO}
\end{figure}

In this paper, we propose a set of hardware-agnostic accelerator orchestration (HALO) principles and a compute-centric message passing interface (C$^2$MPI) specification for enabling host code performance portability. An implementation of the HALO principles can provide a unified orchestration framework to actuate hardware-specific kernels within performance-optimal runtimes and programming models. The proposed HALO principles impose a clear demarcation between hardware-specific and hardware-agnostic software development to allow DMEs and HMEs to work independently in completely decoupled development regions (see Figure \ref{Figure: HALO}) to significantly improve practicability, productivity, and efficiency. To accomplish this, DMEs are restricted to conditioning and steering (orchestrating) data in and out of functional abstractions of hardware-optimized kernels. The hardware-agnostic abstraction of kernels in this regard can be defined by a label and its inputs, outputs, and state variables. Such a functional approach to hardware-agnostic programming is the key to the clear division between the responsibility of DMEs and HMEs. As a result, HMEs will focus on optimizing hardware-specific kernel implementations in their optimal programming environments while being able to eliminate the adoption barrier by leveraging the HALO framework via a unified hardware-agnostic accelerator interface. Furthermore, DMEs will focus on application or algorithm development while being able to maintain a single code flow and effortlessly reap the performance benefits of new hardware accelerators by leveraging the HALO framework via a unified hardware-agnostic application interface. 

The proposed C$^2$MPI specification defines both of the two interfaces and is a compute-centric extension to the traditional MPI specification. It builds upon the MPI paradigm by adding notions of heterogeneous ranks in a parent-child relationship and a distributed remote procedure call (DRPC) execution model for simplicity and ease of adoption. As MPI is the de facto standard in HPC for scaling out (sometimes also up) applications, we chose to extend MPI into the heterogeneous accelerator space as a means to provide existing applications with a natural path for scaling out and up across heterogeneous accelerators. The C$^2$MPI specification not only unifies data computation and communication but also provides unified interfaces for enabling hardware-agnostic programming and transparent interoperability between accelerator resources. Due to the page limitation, this paper focuses on addressing the performance portability perspective of the proposed solution. Therefore, our approaches and their evaluation are primarily presented based on a single host node in the context of enabling host code performance portability. The scalability (out and up) and interoperability perspectives based on multiple host nodes will be addressed in future work. 

In addition, we present HALO 1.0, an open-ended extensible multi-agent software framework that implements the proposed HALO principles and C$^2$MPI specification for enabling the orchestration of performance-optimized, hardware-specific kernels while maintaining hardware-agnostic, performance-portable application host codes across heterogeneous computing devices. The currently supported computing devices are Intel Xeon CPUs, Intel FPGAs, and NVIDIA GPUs. HALO 1.0 consists of two system agents, \textit{i.e.} runtime agent and virtualization agent, which work asynchronously in a star topology. The runtime agent is responsible for implementing and offering the C$^2$MPI, as well as being the crossbar switch for application processes and virtualization agents. The runtime agent also manages system resources, including device buffers, accelerator manifests, kernels, etc. A virtualization agent provides an asynchronous peer that encapsulates hardware-specific compilers, libraries, runtimes, and drivers. The runtime and virtualization agents implement common interprocess communication (IPC) channels for interoperability between multiple virtualization agents, which allows us to scale the number of accelerator types supported while maintaining the simplicity and structure of the framework. 

\noindent The contributions of this work are summarized as follows:
\begin{itemize}
\item We propose the HALO principles for developing hardware-agnostic specifications, programming models, and software frameworks.
\item We propose C$^2$MPI, a compute-centric extension to the traditional MPI specification with newly introduced concepts of heterogeneous ranks to unify the programming and communication interfaces to both HPC applications and accelerators. 
\item We present HALO 1.0, an open-ended extensible multi-agent software framework for enabling the hardware-agnostic programming and performance\nobreakdash-portable execution of a host application across heterogeneous computing devices with great simplicity and extensibility. To the best of our knowledge, this is the first work that aims at enabling hardware-agnostic programming of host applications with true performance portability guarantee while maximizing the reusability of kernel codes across all existing and future accelerator technologies in the context of heterogeneous HPC.
\item We formally define the metric of performance portability score for quantitatively assessing performance portability.
\item We evaluate HALO 1.0 for accelerating eight widely used HPC subroutines based on Intel Xeon E5-2620 v4 CPUs, Intel Arria 10 GX FPGAs, and NVIDIA GeForce RTX 2080 Ti GPUs. The experiment results show that HALO 1.0 allows the same hardware-agnostic application codes, without any change, to run across all the computing devices and achieves the same level of performance compared to the respective hardware-optimized implementations. The kernel performance in HALO 1.0 is equivalent to a consistently top performance portability score of 1.0, which is 2x-861,883x higher than the OpenCL-based solution that suffers from unstably low performance portability scores of 1.2e-6 to 0.62.
\end{itemize}

\section{ \label{sec: background}Background and Related Work}

\subsection{Definitions}
\begin{itemize}
    \item \textit{Programming Model:} The programming methodology used to construct a unified execution flow at the host code level using HALO.
    \item \textit{Portability:} The degree of hardware independence (\textit{i.e.} hardware-independent code flows) an application code binary embeds. 
    \item \textit{Performance Portability}: The degree to which a fully portable host code can maintain the high performance of a hardware-optimized implementation on an accelerator.
    \item \textit{Agent:} An asynchronous process running on an operating system that takes its input from an inter-process communication channel (\textit{i.e.}, a forked process).
    \item \textit{DME} : An application developer or scientist that focuses on conditioning and steering data for pre-defined processing and analytics for scientific discovery.
    \item \textit{HME} : An HPC optimization and/or hardware expert that focuses on developing performance-critical, hardware-optimized device kernels for the data processing and analytics subroutines needed by DMEs.
\end{itemize}

\subsection{\label{subsec: programming model}Heterogeneous Programming Models}
OpenCL, SyCL, OpenMP, OpenACC, and CUDA are programming models for writing programs that execute across heterogeneous computing devices \cite{10.1145/3388333.3388658,10.1145/1229428.1229482,10.1109/WACCPD.2014.10,10.1109/ScalA.2014.8,10.1145/1281500.1281650,10.1145/2807591.2807621,10.1145/1281500.1281647}. But, these programming models are hardly hardware-agnostic as they almost always require custom control flow for each type of accelerator. Developers still need to embed certain hardware-specific details, such as tag dispatch, kernel implementation without abstraction, compute units, work items, and device memory managers, directly into the application codes for optimal performance.  Thus, such hardware-specific optimization still causes code flow divergence and requires DMEs to refactor HPC codes to fit the characteristics of the underlying hardware. Furthermore, these programming models are mostly designed to provide functional portability rather than performance portability.
Although these programming models are not inherently performance-portable, some of them can be practiced in a function-portable fashion, unlike the framework-based solutions mentioned below.

\subsection{\label{subsec: frameworks}Frameworks and Domain-specific Libraries and Languages}
There are various frameworks that provide portability across programming models such as KOKKOS, Thrust, HipSyCL, RAJA. These frameworks give the user an extra level of abstraction above any single programming model interface to enable a semi-unified programming model while providing limited functional and performance portability \cite{10.1145/3388333.3388658, 10.1145/3299771.3299773, 10.1109/LLVM-HPC.2014.9, 10.1145/3302516.3307350, 10.1145/2688500.2688505, 10.5555/3018814.3018822, 10.1145/3293883.3302577,10.1145/3085158.3086159}. 

However, these unifying frameworks suffer from the same pitfalls as their constituent parts. Specifically, these frameworks lack one or more critical components required for providing hardware agnosticism, transparent interoperability, simplicity, and/or extensibility. Such critical components include a complete hardware-agnostic interface for enabling a unified control flow for host codes across all target platforms, transparent interoperability between accelerators, and an open-ended extensible software architecture for allowing the plug-and-play of various types of kernels and accelerators. Different from these solutions, HALO 1.0 seamlessly integrates all these components to fundamentally address the unmet needs of reducing the hardware dependency throughout the application, allowing for hardware-agnostic programming and performance portability of the host code as well as isolating hardware-specific kernels to be developed independently from and invoked uniformly by the host application.

There are also various domain-specific libraries and languages (such as MKL, openBLAS, MAGMA, YASK, Arbiter, Puffin) that provide a great source for popular hardware-optimized software for HPC \cite{10.5555/3241639.3241644, 10.1145/2664666.2664667,10.1145/2661136.2661159, 10.1145/2889420.2889421,10.1145/3375627.3375858,10.5555/3019129.3019133, 10.1145/2830018.2830021,10.5555/2691166.2691168,10.5555/2691166.2691171}. Unfortunately, although these kernels are the most performance-wise reusable code, they are far from portable or extensible. Each library deployment often targets specific hardware, vendor, and/or have a domain-centric interface.

\subsection{\label{subsec: MPI spec}Message Passing Interface (MPI) specification}
MPI defines a standard interface for data movement. MPI has been developed, extended, and refined from 1993 to the present by various organizations, including academic researchers (\textit{e.g.}, ANL, LLNL, University of Tennessee, Cornell University, and University of Edinburgh), library developers (\textit{e.g.}, IBM, Intel, and Cray), and application developers across the HPC spectrum. MPI defines a robust set of interfaces for allocating, sending, and receiving data from CPU processes and exotically from GPUs only. Legacy MPI does not support interoperability of heterogeneous accelerators nor have a kernel execution model for the aforementioned accelerators \cite{10.1145/3295500.3356176,10.1145/2966884.2966894,10.1145/2488551.2488556,10.1145/2488551.2488564,10.5555/3291656.3291696,bruck1997efficient,10.1145/3295500.3356176}. MPI is the de facto standard for scaling out HPC applications across multiple nodes in a cluster environment and earlier to scale up as well.  

The proposed C$^2$MPI specification extends the capabilities of MPI from a communication standard primarily for CPUs to a compute-centric one for heterogeneous accelerators with support for hardware-agnostic programming and execution models. C$^2$MPI introduces a heterogeneous ranking system and a kernel execution model that enables developers to claim and invoke accelerator resources as an abstracted function-specific subroutine. We purposely design C$^2$MPI as an extension of the legacy MPI to simplify and ease the adoption of HALO into existing MPI-enabled applications and minimize the learning curve for developers by keeping with the MPI nomenclature. We also chose MPI in preparation for future work in addressing the scalability and interoperability of accelerators across multiple host nodes. C$^2$MPI seizes on the notion of ranks and introduces heterogeneous ranks to represent accelerator resources. Leveraging C$^2$MPI, HALO inherits the coherency, synchronization, and caching semantics from the legacy MPI.  

\subsection{\label{subsec: RPC}Remote Procedure Calls}
Remote procedure call (RPC) is a protocol commonly used in a client-server system where clients can offload tasks to a remote entity (server). It is a great way to distribute tasks among remote servers and/or peers. They are widely used in web browsers, cloud services (\textit{e.g.}, gRPC) \cite{10.1145/155870.155881}, software-as-a-service platforms, container orchestration (\textit{e.g.}, Kubernetes)\cite{10.5555/3172795.3172840}, massively distributed databases (\textit{e.g.}, Oracle and SAP), high-performance parallel computing offload programming models, and libraries mentioned in Section \ref{subsec: frameworks}. Typically, RPC-based software frameworks (\textit{e.g.}, Azure Databricks, Google AI platform, Apache swarm)\cite{10.1145/3357223.3365870} are used to provide an interface to clients to issue payload, command pairs and have them executed remotely. Similarly, HALO 1.0 leverages the RPC protocol to encapsulate and remotely execute kernels among a network of agent software processes in a peer-to-peer manner.   

\section{\label{sec: HALO Principles}HALO Principles}
HALO principles are the principles to keep in mind when developing hardware-agnostic specifications, programming models, and frameworks. The hallmarks of a hardware-agnostic system are to maintain an interface definition devoid of any vendor-specific, hardware-specific, or computational-task-specific implementations or naming conventions. Interfaces must also be domain-agnostic such that method signatures do not imply functionality but a delivery vehicle. For instance, a method called "execute( kernel1, parameter 1...N)" is domain-agnostic, however, "kernel1(parameter1...N)" is not. Additionally, hardware-agnostic and hardware-specific regions must be clearly defined and decoupled with a robust interoperation protocol. Lastly, abstract functionality must be inclusive of procedures that operate on data and change state. Being domain-agnostic will allow for enormous flexibility and extensibility required to maintain an open-ended HALO software architecture, where HMEs can easily extend the overall system with new accelerator devices and kernel implementations.
The purposes of a hardware-agnostic programming model are twofold. The first is to minimize the amount of hardware-dependent software in a codebase while maximizing the portability of the host code across heterogeneous computing devices. The second is to clearly separate the functional (non-performance-critical) and computational (performance-critical) aspects of the application to simplify the adoption of new accelerator hardware and the development and integration of hardware-specific and hardware-optimized interfaces/kernels.  

\section{\label{sec: C2MPI}C$^2$MPI Specification (Version 1.0)}
\subsection{Overview}
C$^2$MPI defines the unified, domain- and hardware-agnostic interfaces for interfacing, marshaling data, allocating local and remote memory, and executing kernels on heterogeneous HPC systems. C$^2$MPI combines two main interface definitions: a unified application interface for DMEs and a unified accelerator interface for HMEs. To realize such an interface, C$^2$MPI must be domain-agnostic, which refers to the interfaces not being specific to any functional aspect of the underlying kernel. As an opposite example, naming an interface method cblas\_gemm is not domain-agnostic. Additionally, C$^2$MPI must also be hardware-agnostic, which refers to the interfaces not being bound to specific hardware. cudaMalloc for memory management, as an opposite example, is not hardware-agnostic. Formally, C$^2$MPI provides a set of interface definitions that application, framework, and accelerator developers can all agree upon to facilitate the allocation of system resources. C$^2$MPI fuses data marshaling and kernel invocation interfaces to allow the developers to allocate and manipulate optimal accelerator resources without embedding hardware-specific optimization into application codes.  

\begin{table}[h]
\footnotesize
\begin{tabular}{|l|}
\hline
\begin{tabular}[c]{@{}l@{}} "host\_list" : {[}\\ \{"host\_name" : "edge-1.cidse.dhcp.asu.edu",   "port" : "8000", \\ "mode" : "ads\_accel", "max\_slots" : "1"\},\\\\ \{"host\_name" : "turing-4.cidse.dhcp.asu.edu", "port" : "8000", \\ "mode" : "ads\_accel", "max\_slots" : "1"\}\\{]},\\				\\ "func\_list" : {[}\\ \{"func\_alias":"MMM","sw\_fid":"12345", "func\_repl":"1", "platform\_id":"rr\_scat"\},\\ \{"func\_alias" : "EWMM", "sw\_fid":"123456","platform\_id" : "rr\_scat"\},\\ \{"func\_alias" : "SMMM",   "sw\_fid" : "1234567",    "platform\_id" : "rr\_scat"\},\\ \{"func\_alias" : "EWMD",   "sw\_fid" : "12345678",   "platform\_id" : "rr\_scat"\},\\ \{"func\_alias" : "VDP",    "sw\_fid" : "123456789",  "platform\_id" : "rr\_scat"\},\\ \{"func\_alias" : "JS",     "sw\_fid" : "123456789A", "platform\_id" : "rr\_scat"\},\\ \{"func\_alias" : "FC",     "sw\_fid" : "123456789B", "platform\_id" : "rr\_scat"\},\\ \{"func\_alias" : "1DCONV", "sw\_fid" : "123456789C", "platform\_id" : "rr\_scat"\}\\{]},\\ 				\\ "platform\_list" : {[}{]}
\end{tabular} \\ \hline
\end{tabular}
\caption{\label{tab: Configuration File}An example of the configuration file.}
\end{table}

C$^2$MPI leverages MPI semantics to enable domain and hardware agnosticism by utilizing a unified interface applicable to all types of accelerators. System resources can be allocated, marshaled, and invoked through the unified interface. Version 1.0 of the specification includes two types of system resources: 1) handles to functional code segments (a.k.a. kernels); 2) buffer and kernel pipeline allocations. Furthermore, since C$^2$MPI adopts and extends the legacy MPI specification and interfaces, careful considerations are taken to integrate compute-centric capabilities while maintaining backward compatibility with function signatures, programming models, and the overall MPI semantics to facilitate a unified programming model at any scale. With legacy MPI in mind, C$^2$MPI unifies communication and computation orchestration between accelerators and general-purpose CPUs through a heterogeneous parent-child ranking system that describes all computation resources as ranks. Parent ranks (PRs) can allocate and manage child ranks (CRs).  The C$^2$MPI specification is defined by two sub-specifications: one for application parent processes and the other for the accelerator parent processes.In a multi-user context, each application will have a unique ID internally generated, with each child rank creation being tagged with the application ID. The HALO system will keep track of which CRs correspond to which application context. 

\subsection{\label{subsec: Parent Ranks}Parent Rank (PR)}
Application PRs live inside the hardware-agnostic region of an application. Application PRs are not guaranteed to be performance-portable. Application PRs are synonymous with traditional MPI ranks. They have the full capabilities of a typical MPI-based rank process along with the added capabilities for CR management. Both application and accelerator processes can allocate and manage CRs.  
Accelerator PRs, in addition to managing their own CRs, are responsible for hardware management, kernel retrieval, registration, and execution and maintaining system resources allocated by application PRs. Similar to MPI-based applications, jobs can instantiate multiple application PRs, and each PR can be multi-threaded, making requests into the child management system asynchronously. Therefore, the C$^2$MPI interfaces are thread-safe. But, version 1.0 of C$^2$MPI does not allow system resources to be shared across the boundary of PRs.  Communicators and groups can be created/defined on PRs similar to traditional MPI ranks. Communicators can be formed on child ranks to utilize collective operations.

\begin{table}[]
\scriptsize
\centering
\begin{tabular}{|l|l|l|l|}
\hline
\rowcolor{gray!20}
\textbf{\begin{tabular}[c]{@{}l@{}}Kernel \\ Attribute\end{tabular}} & \textbf{Description} & \textbf{\begin{tabular}[c]{@{}l@{}}Kernel \\ Attribute\end{tabular}} & \textbf{Description} \\ \hline
VID & HW Vendor ID & SW\_PID & SW Product ID \\ \hline
PID & HW Product ID & SW\_VID & SW Vendor ID \\ \hline
SS\_VID & HW Sub-system Vendor ID & SW\_FID & SW Function ID \\ \hline
SS\_PID & HW Product ID & SW\_VERID & SW Version ID \\ \hline
\end{tabular}
\caption{\label{table: kernel attributes} Kernel attributes for kernel selection process.}
\end{table}

\subsection{\label{subsec: child rank}Child Rank (CR)}
CRs are the virtual abstraction of a system resource in the form of an opaque handle, similar to a PR but with limited capabilities. Such a system resource is not inherently tied to any physical resource at runtime, and the runtime agent has full authority to move both functionality and allocation to compatible accelerators on the network while assuring computation integrity. CRs can be allocated via an application or accelerator PR, with both having the life span of the job issuing requests. CRs can represent a single resource or a set of resources in parallel or pipeline. A pipeline of resources is a series of dependent kernel invocations. CRs can be deallocated via C$^2$MPI interfaces, and the resources are freed when \textit{MPIX\_Finalize} gets executed.  

\begin{table}[h]
\scriptsize
\centering
\begin{tabular}{|l|}
\hline
\rowcolor{gray!20}
\textbf{Function Signature} \\ \hline
MPIX\_Send (const void *,  int, MPIX\_Datatype,  int, int, MPIX\_Comm) \\ \hline
\rowcolor{gray!20}
\textbf{Invocation Example} \\ \hline
MPIX\_Send (CompObj, 1, MPIX\_CompObj, 20, 0, comm) \\ \hline
\end{tabular}
\caption{ \label{tab: MPIX Send function signature} The \textit{MPIX\_Send} function signature (left) and invocation example (right).}
\end{table}

CRs have attributes that can be defined statically from a configuration file or dynamically at runtime. Refer to Table \ref{table: kernel attributes} for the attribute list and the description for the resource selection process. Developers can create an alias for resources and specify kernels and optional executors via a configuration file defining kernel attributes. CR management uses these attributes to allocate resources according to the resource alias. We merge the legacy MPI host file with the accelerator manifest to provide a unified configuration file for end-users (see Table \ref{tab: Configuration File}).  

The configuration file is broken into three sections. The first section is the host list, which is synonymous with the MPI host file. The second section is the resource list for CR definitions. The third section contains system configuration details, which configures the hardware recommendation strategy for allocating resources. Finally, each invokable resource must either have a functional or class identification (sw\_fid, sw\_clid) in the configuration file as these IDs are the main mechanism for resource lookup. If the function or class identifier cannot be located in the runtime agent manifest, the user-defined function will execute in a fail-safe mode to assure system resilience and maintain functional portability for that specific kernel.
 
\subsection{Unified Compute-object Structure and Enumerations}
The compute-object and its associated enumerations are the primary vehicles for generalizing and encapsulating all arguments to construct complex RPCs (see Figure \ref{Figure:UML}). "Complex" refers to the kernel invocations that have multiple applications and system resources associated with the request.  
\textit{MPIX\_ComputeObj} implements a reflective pattern to encapsulate and extend the type erasure techniques used in legacy MPI. The unified compute object makes marshaling straightforward when communicating RPCs among PRs. The enumerations (\textit{MPIX\_TYPES}) are used to differentiate between internal and external buffers. Internal buffers are managed by the HALO framework, and external buffers are managed by the PRs. Both buffers can also persist across multiple kernel invocations represented in the compute-object by a handle, be allocated by \textit{MPIX\_CreateBuffer}, and be described with an enumeration prefixed by internal buffers (see Figure \ref{Figure:UML}). Compute-objects that use only external buffers are considered stateless RPC invocations, and those with internal buffers are considered stateful RPC invocations. To allow one to utilize unified memory to minimize communication overhead, we introduce the MPIX variance of MPI\_Alloc\_mem. The use of unified memory eliminates the internal copy of data buffers similar to the intra-node communication in legacy MPI. 

\subsection{Data-Movement Interface}
The backbone of C$^2$MPI is the point-to-point communication between parent and CRs using \textit{MPIX\_Send} and \textit{MPIX\_Recv} methods. These methods maintain the signature of their legacy MPI counterparts (see Table \ref{tab: MPIX Send function signature}). The compute-centric extension of these methods comes into perspective during the invocation when the unified compute-object, enumerations, and target CRs are applied (see Table \ref{tab: MPIX Send function signature}). The compute-object can be marshaled to and from child and PRs via these two methods. 

\begin{figure}[t]
\centering
  \includegraphics[scale=.45]{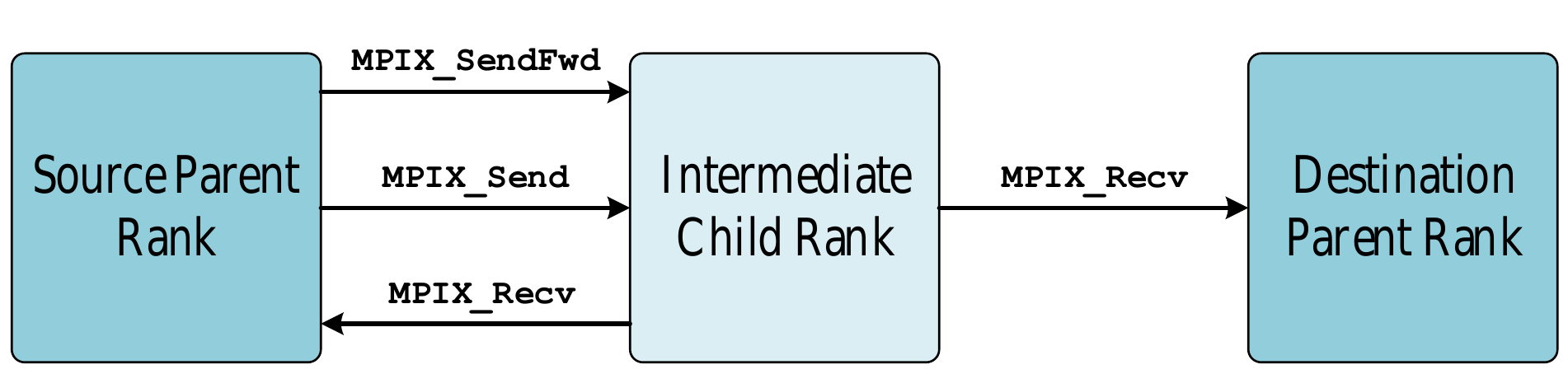}
  \caption{The message flow for \textit{MPIX\_Send}, \textit{MPIX\_Recv}, and \textit{MPIX\_SendFwd}.}
  \label{Figure: MPIX_send message flow}
\end{figure}

\begin{table}[h]
\scriptsize
\centering
\begin{tabular}{|l|}
\hline
\begin{tabular}[c]{@{}l@{}}int MPIX\_Claim (func\_alias, failsafe\_func, overrides, child\_rank);\\ int MPIX\_CreateBuffer (child\_rank, is\_signed, is\_float, type\_size, \\                                        vector\_size, child\_rank\_buf);\\ int MPIX\_Free (child\_rank);\end{tabular} \\ \hline
\end{tabular}
 \caption{\label{tab: C2MPI allocation-deallocation}Allocation/deallocation prototypes for C$^2$MPI.}
\end{table}

When using an \textit{MPIX\_Send}, the resulting compute-object returns to the source PR by default. To forward the compute-object to a different PR, one can use the \textit{MPIX\_SendFwd} method. The \textit{MPIX\_SendFwd} method looks similar to the \textit{MPIX\_Send} interface with an added parameter for the destination PR (see Figure \ref{Figure: MPIX_send message flow}). Note that one can bypass using the unified compute-object when RPCs are simple structures or single array buffers and pass payloads as one would with traditional MPI. The single-input optimization saves the step of encapsulating a multi-input payload. 
Similar to MPI, the tagging mechanism can be used to retrieve results from multiple \textit{MPIX\_Send} out of order or from different threads. Transmissions can be set with a tag such that a \textit{MPIX\_Recv} can be used to retrieve data out of order. Repeated calls to \textit{MPIX\_Recv} with the same tag will result in a FIFO behavior. 

\subsection{Resource Allocation/Deallocation Interface}
There are three methods to support the allocation of virtual accelerators resources: \textit{MPIX\_Claim}, \textit{MPIX\_CreateBuffer}, and \textit{MPIX\_Free} (see Table \ref{tab: C2MPI allocation-deallocation}). The \textit{MPIX\_Claim} interface takes in an alias tag that references an entry in the configuration file's func\_list (see Table \ref{tab: Configuration File}) that describes the functionality of the virtual resource (a CR) using the kernel attributes in Table \ref{table: kernel attributes}. These parameters can be overwritten at runtime through the \textit{MPI\_Info} argument. In addition, \textit{MPIX\_Claim} accepts a \textit{fail\_safe} callback that takes in a unified compute object and outputs a unified compute-object in case no accelerator resources are available. \textit{MPIX\_Claim} returns a status and a handle to the CR used in the data movement interface. These CRs are stateless by default.

\textit{MPIX\_CreateBuffer} is an interface to allocate internal memory. This resource can be associated with the HALO framework or other CRs. Passing zero as a CR handle will allow the method to associate the memory allocation to the HALO framework. \textit{MPIX\_CreateBuffer} is the main vehicle for creating a state from a stateless CR, created by \textit{MPIX\_Claim}. Finally, \textit{MPIX\_Free} takes in a CR, deallocates the resource, and returns null as a handle.

\begin{figure}[h]

 \includegraphics[scale=0.55]{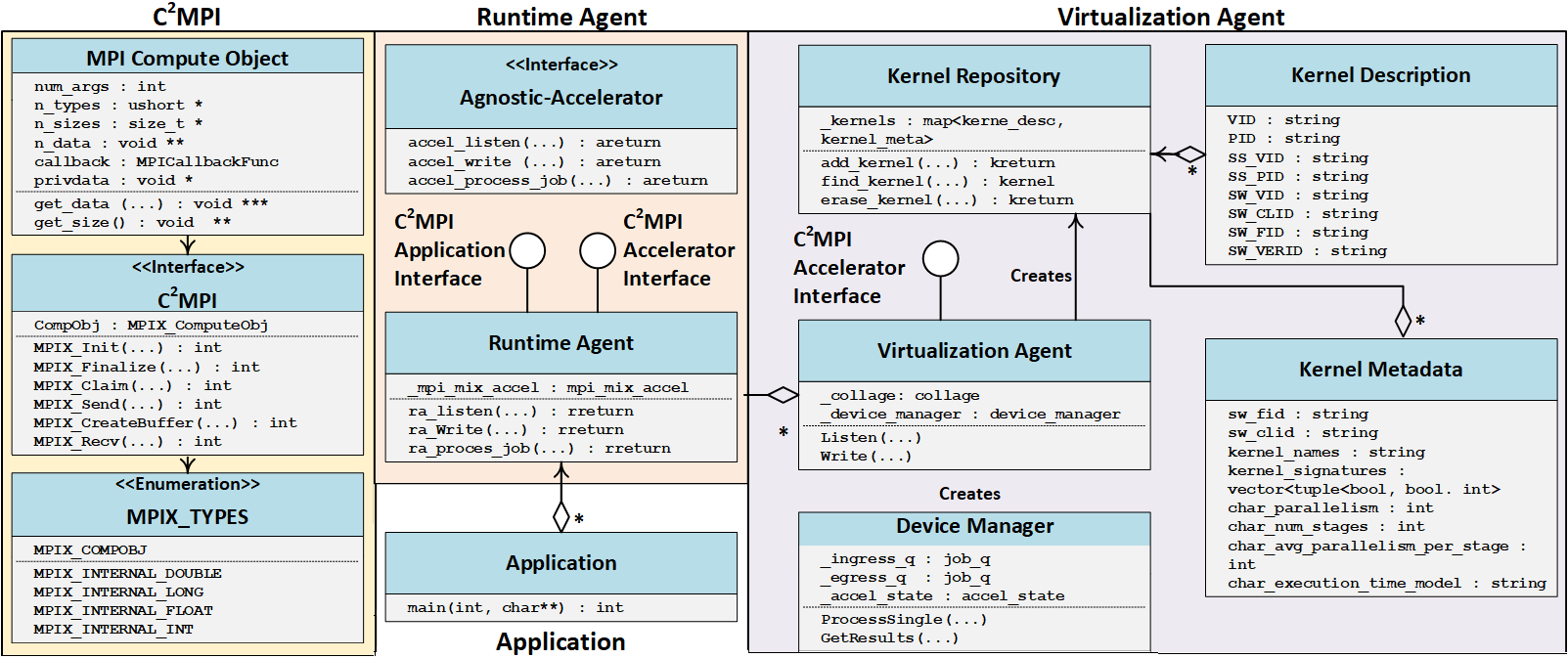}
  \caption{A unified modeling language diagram of the HALO 1.0 software stack.}
  \label{Figure:UML}
\vspace{-1.5em}
\end{figure}

\section{Framework (HALO 1.0) Design}
\subsection {Framework Concepts}

\subsubsection{Domain/Hardware-Agnostic Application Interface}
HALO 1.0 implements the application subset of C$^2$MPI in the C language to offer a domain- and hardware-agnostic interface to DMEs (application developers, scientists) to enable performance portability. The C$^2$MPI application interface will allow DMEs to abstract hardware details from application codes while orchestrating hardware-optimized kernels and achieving the best-in-class performance of accelerator resources without needing to modify application code at all. The  C$^2$MPI application interface allows for DMEs to harden the performance portability of application codes for both the current and future accelerator technologies. 

\subsubsection{Domain-Agnostic Accelerator Interface}
Similar to the C$^2$MPI application interface, HALO 1.0 implements the C$^2$MPI accelerator interface in the C language to provide a domain- and hardware-agnostic interface for HMEs with transparent interoperability between system resources and the runtime and virtualization agents. Additionally, the C$^2$MPI accelerator interface provides the specification to support DRPC, reading meta-data, manifesting, and other capabilities provided by remote system resources, including system performance, rank requests, and host/device heap memory allocation.  

\subsubsection{Agent Interoperability}
\label{subsec: Agent interoperability} 
HALO 1.0 enables interoperability among multiple virtualization agents, each supporting a different accelerator device and PRs via the runtime agent. The interoperability stems from a combination of purely asynchronous message protocols and specifications based on which an accelerator resource can leverage other system resources, such as runtime and virtualization agents and system-wide memory resources. The interoperability protocol utilizes an asynchronous request-and-response protocol allowing for multiple messages from various runtime and virtualization agents to be serviced in parallel. The accelerator interface and interoperability protocol are implemented in a pipeline (or a chain of responsibility pattern) in the virtualization agent to support code reuse and modularization. 
\subsubsection{Accelerator Multi-source Kernels Repository}
The HALO framework employs a multi-source approach to enable hardware-agnostic programming. Hardware-specific kernels are placed in separate source files that are compiled/linked dynamically by the virtualization agent. The indexing of these kernels is governed by the lookup system discussed in Section \ref{subsec: child rank}.
The kernel specifications, such as compatible execution hardware, function signatures, and other meta-data for each kernel, are embedded and communicated to the virtualization agent dynamically or statically. The kernels have a special hardware-agnostic extension (*.ha) that comprises both the kernel specification and binaries.   

\subsubsection{Multi-Agent System}
HALO 1.0 is a multi-agent system. The multi-agent structure is a direct response to the HALO principles for providing an open-ended extendable architecture that allows runtime and virtualization agents supporting various devices and system functionalities to be dynamically connected and disconnected from the runtime agents without affecting hardware-agnostic applications. The plug-and-play nature of the HALO 1.0 architecture, along with clear interface definitions, will allow HMEs to develop, evaluate, and deploy HALO-compatible, hardware-optimized kernels rapidly.

\subsubsection{Hierarchical Parallelism}
HALO 1.0 manages parallelism at multiple levels, with the first being the parallelism in the Parent Ranks (PR), which are synonymous to the parallelism obtained by launching multiple legacy MPI ranks. The next level of parallelism comes from the Child Ranks(CR). CRs extend the principle of legacy MPI to accelerators by linking a functional description and a logical accelerator context driven by the configuration file. The Runtime Agent (RA) brokers which virtualization agent should manage jobs; then, the virtualization agent's multi-threaded frontend pipeline feeds a device manager that manages device (accelerator) level parallelism (ex. CUDA streams, async kernel launches). The application can inherit first level parallelism by creating multiple CRs within the same process(s) or thread(s) and executing within process or thread runtime. Concurrent execution is similar to launching multiple thread-safe asynchronous kernel launches with the C$^2$MPI. Multiple MPIX\_* can be launched across multiple threads to leverage the HALO parallelism.

\begin{figure}[h]
\centering
  \includegraphics[scale=0.8]{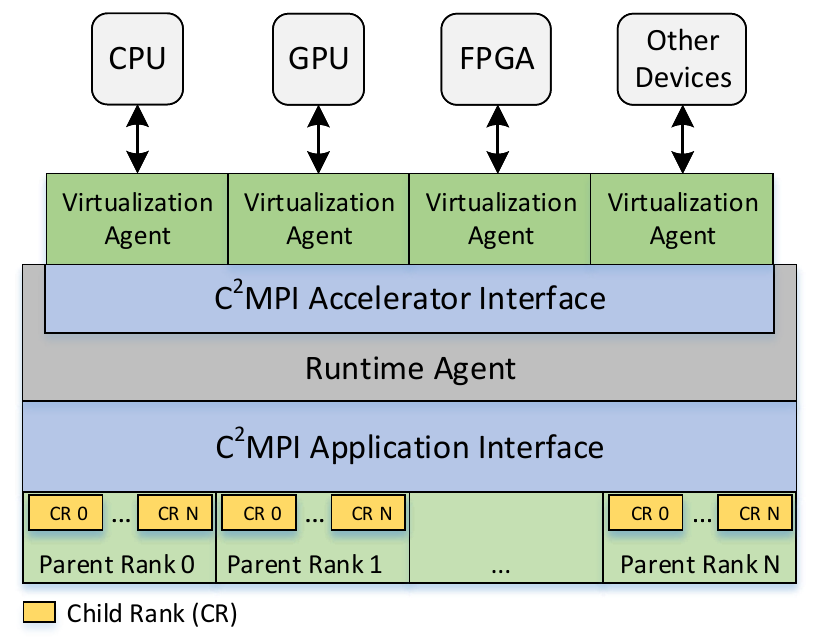}
  \caption{A high-level software stack diagram of HALO 1.0.}
  \label{Figure:OA}
\end{figure}

\subsection{Software Architecture Overview}
HALO 1.0 is a multi-agent C/C++ software framework that provides hardware-agnostic accelerator orchestration by implementing the C$^2$MPI specification for both application and accelerator frontends (see Figure \ref{Figure:OA}). The backend implements a C++-based multi-agent system that includes runtime agents and virtualization agents working together asynchronously. Each virtualization agent implements a different device-specific runtime for kernels to execute in, while the runtime agent implements both the C$^2$MPI application and accelerator interfaces (see Figure \ref{Figure:UML}). Topologically, the system is built on a star pattern where the runtime agent acts as a bridge between different ranks and virtualization agents. All the agents communicate through a unified memory, using ZeroMQ to pass only pointers to shared memory. This keeps the overhead of intra-HALO message passing invariant to the working set size (WSS). Both runtime agent and virtualization agents present their physical resources in a virtual manner. This allows the agents to redirect requests dynamically. Each agent includes facilities to track resource allocation.

The runtime and virtualization agents are loosely coupled and interconnected by a domain-agnostic protocol (see Table \ref{tab: C2MPI allocation-deallocation}). These agents operate completely asynchronously with each other. The application PRs operate synchronously or asynchronously to and from the runtime agent. In order to protect against system-wide race conditions and deadlocks, synchronization points only occur at the application PR thread level. Therefore, if a thread in an application PR calls a blocking C$^2$MPI method,  the blocking mechanism will only block locally to that thread.

There are currently five virtualization agents developed in HALO 1.0: one for Intel CPU-based libraries, Intel CPU OpenCL/SyCL runtimes, NVIDIA CUDA/Thrust runtimes, NVIDIA GPU OpenCL runtimes, and Intel FPGA OpenCL runtimes, respectively. Each virtualization agent is interconnected via a domain-agnostic interoperability protocol using ZeroMQ IPC to transfer information between the runtime and virtualization agents (see Sections \ref{subsec: Agent interoperability} and \ref{subsec: VA}).

\subsection{\label{subsec: RA}Runtime Agent}
\begin{figure}[t]
  \centering
  \includegraphics[scale=0.6]{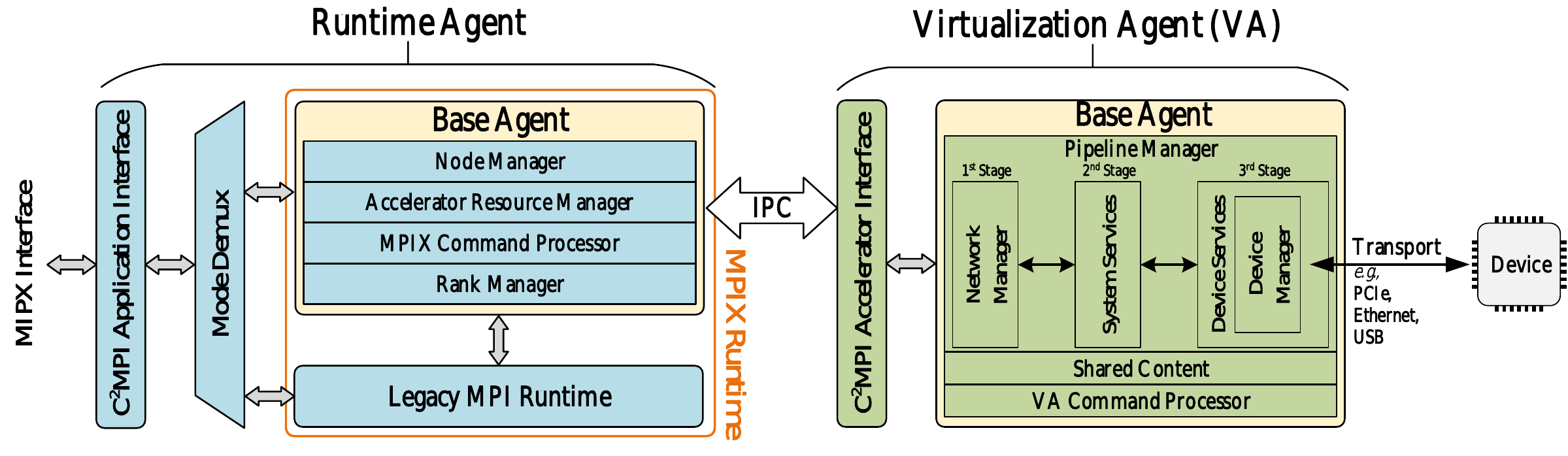}
  \caption{A low-level software stack diagram of HALO 1.0.}
  \label{Figure: Runtime and Virtualization Agents} 
\end{figure}

There is one runtime agent process for each parent rank in progress, providing multi-tenancy support. The runtime agent is a duo-thread agent representing the C$^2$MPI application and accelerator interfaces. The first thread shares the same virtual memory space as the application. The application library frontend interfaces with the MPIX runtime via the two unidirectional lock-free queues, interconnecting the first and second thread bidirectionally. The first thread is a thin thread that handles the synchronicity requirements for the interface without burdening the MPIX runtime. When calls are made to the MPIX runtime, the mode demultiplexer determines whether it takes a native MPIX runtime or legacy MPI runtime route (see Figure \ref{Figure: Runtime and Virtualization Agents}). When the host code invokes multiple kernels, the runtime agent uses a round-robin recommendation strategy to assign a candidate accelerator resource. This functionality will be evaluated in the forthcoming study on scalability based on multiple host nodes.

The second thread models a proactor pattern (via ZeroMQ IPC channels) that manages interoperability between the MPIX and Legacy MPI runtimes. The second thread takes messages from the application and both runtime and virtualization agents and processes them via the command processor. Furthermore, The second thread manages system resources, converts synchronous messages to asynchronous messages, encapsulates, serializes, and deserializes messages in/out of the agent interoperability protocol. The runtime agent bridges the system messages between MPIX and MPI runtimes asynchronously. The MPIX and MPI runtimes are synchronized through two queues that feed the second thread.

\subsection{\label{subsec: VA} Virtualization Agent }
The virtualization agent provides an asynchronous peer that encapsulates hardware-specific compilers, libraries, runtimes, and/or drivers.
The virtualization agent implements a chain of responsibility pattern, with the frontend being a proactor enabled by ZeroMQ IPC channels. The virtualization agent embodies a three-thread, three-stage pipeline (see Figure \ref{Figure: Runtime and Virtualization Agents}), where each stage is asynchronous and interconnected via lock-free queues. This architecture is critical to maximizing code reuse when refactoring HPC applications for various accelerators. The first stage is a network manager that deserializes and converts messages between the interconnect protocol format and an object-oriented format. The network manager also places these objects into a shared memory content store to eliminate copies and have a central point to recall messages to be handled by the different stages. The lock-free queues pass around references to the shared memory content store in the form of transaction chains of the transaction ID.
The second stage is the system services, which manage requests that are resolvable without hardware intervention. The system services include stored data from the third stage as well as kernel manifest hardware specifications and runtime metrics. 
The third stage is the device services, where the encapsulation of the vendor logic first occurs. The device services handle device-specific details and integrate the device manager into the virtualization pipeline. Furthermore, the device services manage the kernel repository and pass them to the device manager along with the input payload. The device manager does all the heavy lifting, such as configuring the device, allocating device memory, loading kernels, and multiplexing (time and space) the devices. The device manager performs the required configurations, data movements, and invocations applicable to the corresponding RPCs for interfacing with a runtime, framework, or library. 

\section{Evaluation}

\subsection {Experiment Setup}
To assess performance portability (independently from scalability), we evaluate HALO 1.0 on a single host node across multiple accelerators independently in comparison with existing solutions. 
We evaluate the performance of four different types of implementations for eight widely used HPC kernels based on three different types of computing devices (\textit{i.e.}, CPU, GPU, FPGA). The four implementation types are hardware-optimized baseline implementations using the best available vendor-suggested libraries or frameworks, hardware-specific and hardware-agnostic implementations using OpenCL, and hardware-agnostic implementations based on HALO 1.0. The eight HPC kernels are matrix-matrix multiplication (MMM), element-wise matrix multiplication (EWMM), sparse matrix-matrix multiplication (SMMM), matrix-vector multiplication (MVM), element-wise matrix division (EWMD), vector dot-product (VDP), Jacobi solver (JS), and one-dimensional convolution (1DConv). These kernels represent a set of core computational workloads prevalent in multiple HPC disciplines, such as computational fluid dynamics, computational material science, weather forecasting, and deep learning. Furthermore, the baselines are taken from well-known hardware-optimized kernels from vendor libraries. Since performance portability is respective to a hardware-optimized implementation on an accelerator, specific kernel optimization and functionality are irrelevant and outside the scope of this study. The kernels are evaluated on a HALO test harness configured with 1 PR and claiming 1 CR, associated with a single accelerator.

The performance metrics used for evaluation are HALO overhead (\(T1\)) between the runtime and virtualization agents (IPC with ZeroMQ), hardware data transfer (offload) time (\(T2\)), kernel execution time (\(T3\)), and the total runtime (\(T4 = T1 + T2 + T3\)). The HALO overhead is the round-trip time from sending a request from the host code to receiving a response minus T2 and T3. In other words, the HALO overhead measures the runtime overhead imposed by the HALO software framework only. Specifically, the HALO overhead is mainly contributed by the IPC overhead and the agent response times. The HALO overhead excludes the time that the device runtime (\textit{e.g.,} CUDA, OpenCL) within the device manager requires to launch kernels and retrieve application data as such overhead is not imposed by HALO.  
It should be noted that the HALO overhead is invariant to the WSS of the kernel since HALO implements a unified memory model and only passes pointers to data. 

The experiments are performed on two server nodes. The GPU node runs on Ubuntu 18.04 and is equipped with 64 GB of DRAM and dual sockets of Intel Xeon E5-2620 v4 CPU, each hosting two NVIDIA GeForce RTX 2080 Ti GPUs. The FPGA node runs on Redhat 7 and is equipped with 32 GB of DRAM and an Intel Xeon E3-1275 v5 CPU hosting two BittWare 385A FPGA accelerators (Intel Arria 10 GX1150 FPGA). The CPUs used for evaluation are the ones on the GPU node. NVIDIA driver 440.100 and CUDA toolkit 10.2 are used for the GPUs. Intel FPGA SDK for OpenCL with Quartus Prime Pro 19.1 is used for the FPGAs. The SyCL DPC++ beta compiler from DPC++ daily 2020-07-23 is used.  

The CPU-optimized baseline implementations leverage Intel Math Kernel Library (MKL) and handwritten C++. The GPU-optimized baseline implementations leverage a combination of Thrust, NVIDIA libraries, and handwritten C++ (including CUDA). The FPGA-optimized baseline implementations are based on OpenCL 1.1 with high-level synthesis. The hardware-specific OpenCL implementations leverage standard OpenCL with hardware-specific optimization, such as SIMD width, compiler flag optimization, memory coalescing, channels extension, and other hardware-specific attribute optimization. The hardware-agnostic OpenCL implementations remove such hardware-specific optimization in the host or device code. It should be noted that these hardware-optimized baselines represent the highest performing implementations. We purposely chose these baselines for comparison to demonstrate that HALO is able to maintain the highest performance achievable by the hardware-optimized implementations while enabling the hardware-agnostic programming of host applications. As the existing portability frameworks \cite{10.1145/3318170.3318193, 10.5555/3314872.3314914, 10.1145/3204919.3204939} or heterogeneous programming languages fail to provide complete host code portability (as remarked in the related work section), they are not considered as counterparts for comparison in this study. Differently, OpenCL codes can be developed in a hardware-agnostic fashion, although often at the expense of performance degradation. Thus, OpenCL is considered as the state-of-the-art viable portability framework and used as the reference for comparison in this study. 

The hardware-agnostic HALO implementations leverage HALO 1.0 configured with a runtime agent and five virtualization agents. The virtualization agents support the virtualization of Intel CPU-based libraries, Intel CPU OpenCL/SyCL, NVIDIA CUDA/Thrust, NVIDIA GPU OpenCL, and Intel FPGA OpenCL runtimes. Taking advantage of hardware agnosticism and transparent interoperability, HALO 1.0 can always leverage hardware-optimized baseline implementations to accelerate the hardware-agnostic HALO implementations. The template of the host source code of a HALO implementation is shown in Table \ref{tab:host}.  

To reveal the performance degradation of transitioning from baseline to OpenCL implementations, we define the metric of performance penalty (\%) as $(T3_{OpenCL}-T3_{Baseline})/{T3_{Baseline}}\times100$.
To compare the performance portability between the two hardware-agnostic solutions, we define the metric of performance portability score as $T3_{Baseline}/T3_{Hardware-agnostic}$. Performance portability score ranges from 0 to 1, which quantifies the ability of a hardware-agnostic implementation to maintain a high performance (low kernel execution time) relative to the hardware-optimized implementation across different computing devices. To reveal the impact of HALO overhead on the total runtime, we define the metric of the HALO overhead ratio as $T1_{HALO}/T4_{HALO}$.
\begin{table}[t]
\scriptsize
\centering
\begin{tabular}{|p{10cm}|}
\hline
\begin{tabular}[c]{@{}l@{}}ulong child\_rank=0;\\ const char * tag\_alias = "kernel\_alias\_HERE";\\ MPIX\_ComputeObj compObj;\\ MPIX\_Initialize(...) ;\\ MPIX\_Claim(tag\_alias, ..., \&child\_rank);\\ MPIX\_Send(compObj, ..., child\_rank, ...); //prepare MPIX\_ComputeObj...\\ MPIX\_Recv(compObj, ..., child\_rank, ...);\\ MPIX\_Finalize(…); //check results...\\ return 0;\end{tabular} \\ \hline
\end{tabular}
\caption{\label{tab:host}A template of HALO 1.0 application codes that are both hardware- and domain-agnostic.}
\end{table}

\begin{table}[h]
\scriptsize
\tabcolsep=0.15cm
\centering
\renewcommand{\arraystretch}{1.3}
\begin{tabular}{|c|c|c|c|c|c|c|}
\hline
\multirow{2}{*}{\textbf{Kernel Name}} &
\multicolumn{3}{c|}{\textbf{HS-OpenCL}} & \multicolumn{3}{c|}{\textbf{HA-OpenCL}} \\ \cline{2-7} 
& \textbf{CPU} & \textbf{GPU} & \textbf{FPGA$^{\#}$} & \textbf{CPU} & \textbf{GPU} & \textbf{FPGA} \\ \hline \hline
MMM & 204\% & 47\% & 0\% & 1,892\% & 2,865\% & 246,479\% \\ \hline
EWMM & 54\% & 58\% & 0\% & 162\% & 131\% & 5.6\% \\ \hline
SMMM & 484\% & 0\% & 0\% & 4,491\% & 897\% & 9,778\% \\ \hline
EWMD & 6.2\% & 46\% & 0\% & 60\% & 129\% & 1.4\% \\ \hline
VDP & 0\% & 3.9\% & 0\% & 349\% & 78\% & 1,157\% \\ \hline
JS & 22\% & 2,430\% & 0\% & 215\% & 3,233\% & 1,440\% \\ \hline
MVM & 63\% & 400\% & 0\% & 357\% & 376,300\% & 3,214\% \\ \hline
1DConv & 51\% & 14\% & 0\% & 182\% & 738,657\% & 58,182\% \\ \hline
\end{tabular}
\begin{tablenotes}
\item[\tnote{\textdagger}] 
$^{\#}$ The HS-OpenCL FPGA implementations are also the FPGA-optimized baselines.
\end{tablenotes}
\caption{\label{table: performance penalty} Performance penalty (\%) of hardware-specific (HS) and hardware-agnostic (HA) OpenCL implementations. Lower is better.}
\end{table}

The experiments are conducted using the WSSs ranging from 48MB to 1GB. It should be noted that T2, and T3 all increase near-linearly as WSS scales up. As a result, we observe from the experiment results that the performance portability score and the HALO overhead are all invariant to WSS, benefiting from the use of unified memory. Therefore, the WSS range used is sufficient to project such invariance for larger WSS as well.

\subsection{Experiment Results} 
Table \ref{table: performance penalty} shows the performance penalty of hardware-specific and hardware-agnostic OpenCL implementations. For the CPU and GPU, the hardware-specific OpenCL implementations suffer from a performance penalty of 0\%-484\% and 0\%-2,430\%, respectively, with the majority achieving $<$63\% performance penalty over the hardware-optimized baselines. 

The performance impact after removing hardware-specific optimization paints a grimmer picture. For the CPU, GPU, and FPGA, the hardware-agnostic OpenCL implementations suffer from a much bigger performance penalty of 60\%-4,491\% and 78\%-738,657\%, and 1.4\%-246,479\%, respectively. Therefore, such a large variance of performance penalty makes OpenCL hardly a practical solution to hardware-agnostic programming for the future of heterogeneous HPC.

\begin{table}[h]
\scriptsize
\tabcolsep=0.08cm
\centering
\renewcommand{\arraystretch}{1.3}
\begin{tabular}{|c|c|c|c|c|c|c|}
\hline
\multirow{2}{*}{\textbf{Kernel Name}} & \multicolumn{3}{c|}{\textbf{HALO (HALO/HA-OpenCL)}} & \multicolumn{3}{c|}{\textbf{HA-OpenCL}} \\ \cline{2-7} 
 & \textbf{CPU} & \textbf{GPU} & \textbf{FPGA} & \textbf{CPU} & \textbf{GPU} & \textbf{FPGA} \\ \hline \hline
MMM & 1.00 (\textbf{20x}) & 1.00 (\textbf{30x}) & 1.00 (\textbf{2,466x}) & 0.05 & 0.03 & 4.06e-4 \\ \hline
EWMM & 1.00 (\textbf{3x}) & 1.00 (\textbf{2x}) & 1.00 (\textbf{4x}) & 0.38 & 0.43 & 0.23 \\ \hline
SMMM & 1.00 (\textbf{46x}) & 1.00 (\textbf{10x}) & 1.00 (\textbf{99x}) & 0.02 & 0.10 & 0.01 \\ \hline
EWMD & 1.00 (\textbf{2x}) & 1.00 (\textbf{2x}) & 1.00 (\textbf{4x}) & 0.62 & 0.44 & 0.27 \\ \hline
VDP & 1.00 (\textbf{4x}) & 1.00 (\textbf{2x}) & 1.00 (\textbf{13x}) & 0.22 & 0.54 & 0.08 \\ \hline
JS & 1.00 (\textbf{3x}) & 1.00 (\textbf{33x}) & 1.00 (\textbf{15x}) & 0.32 & 0.03 & 0.06 \\ \hline
MVM & 1.00 (\textbf{5x}) & 1.00 (\textbf{94,100x}) & 1.00 (\textbf{33x}) & 0.22 & 1.10e-5 & 0.03 \\ \hline
1DConv & 1.00 (\textbf{3x}) & 1.00 (\textbf{861,883x}) & 1.00 (\textbf{581x}) & 0.36 & 1.20e-6 & 1.72e-3 \\ \hline
\end{tabular}
\begin{tablenotes}
\item[\tnote{\textdagger}] 
HA = Hardware-agnostic
\end{tablenotes}
\caption{\label{table: portability scores} Performance portability score comparison between HALO and hardware-agnostic OpenCL.}
\end{table}

Although not represented, the best baseline performance for FPGAs should be from RTL-based implementations. Due to time constraints, the hardware-specific OpenCL implementations are also used to represent the FPGA-optimized baseline; hence no degradation is reported in the FPGA column. However, the most to suffer due to the transition from a baseline to a hardware-agnostic OpenCL implementation are FPGAs. The reason is that the implementations fail to explore spatial or temporal parallelism in FPGAs without hardware-specific optimization. Similarly, due to all three accelerators evaluated do not have just-in-time (JIT) compilers, we restricted the use of runtime recompilation and the use of the associated compiler flags for optimization purposes from the hardware-agnostic implementations causing further performance drops.

\begin{table}[h]
\scriptsize
\tabcolsep=0.13cm
\centering
\renewcommand{\arraystretch}{1.3}
\begin{tabular}{|c|c|c|c|c|c|c|c|}
\hline
\multirow{3}{*}{\textbf{Kernel Name}} & \multicolumn{4}{c|}{\textbf{Time (ms)}} & \multicolumn{3}{c|}{\multirow{2}{*}{\textbf{\begin{tabular}[c]{@{}c@{}} HALO Overhead Ratio \\ (T1/T4)\end{tabular}}}} \\ \cline{2-5}
 & \multirow{2}{*}{\textbf{T1}} & \multicolumn{3}{c|}{\textbf{T4}} & \multicolumn{3}{c|}{} \\ \cline{3-8} 
 &  & \textbf{CPU} & \textbf{GPU} & \textbf{FPGA} & \textbf{CPU} & \textbf{GPU} & \textbf{FPGA} \\ \hline \hline
SMMM & 0.0019 & 43 & 50 & 1843 & 0.0044\% & 0.0038\% & 0.0001\% \\ \hline
MMM & 0.0019 & 362 & 124 & 676 & 0.0005\% & 0.0015\% & 0.0003\% \\ \hline
VDP & 0.0019 & 234 & 151 & 131 & 0.0008\% & 0.0013\% & 0.0015\% \\ \hline
EWMM & 0.0019 & 196 & 254 & 729 & 0.0010\% & 0.0007\% & 0.0003\% \\ \hline
EWMD & 0.0019 & 306 & 258 & 718 & 0.0006\% & 0.0007\% & 0.0003\% \\ \hline
JS & 0.0019 & 952 & 321 & 4112 & 0.0002\% & 0.0006\% & 0.0000\% \\ \hline
1D Conv & 0.0019 & 1654 & 447 & 490 & 0.0001\% & 0.0004\% & 0.0004\% \\ \hline
MVM & 0.0019 & 63 & 318 & 274 & 0.0030\% & 0.0006\% & 0.0007\% \\ \hline

\end{tabular}
\begin{tablenotes}
\item[\tnote{\textdagger}] 
T1 = HALO (SW) Overhead (Agent-Agent Data Transfer Time), T4 = Total Runtime.
\end{tablenotes}
\caption{\label{table: sw_overhead} Software overhead of HALO 1.0.}
\end{table}
\raggedbottom

Table \ref{table: portability scores} shows the performance portability score comparison between the HALO and hardware-agnostic OpenCL implementations. The hardware-specific OpenCL implementations are not portable at all in our definition, thus have no performance portability score. Because HALO 1.0 adds a lean level of abstraction and features transparent interoperability and extensibility, the virtualization agents can always leverage hardware-optimized baseline implementations to accelerate the hardware-agnostic application codes. With the abstraction layer provided by HALO 1.0, we can hide all hardware-specific implementation details that are critical for performance from the hardware-agnostic applications while still assuring the same performance as the hardware-optimized baselines across the board. Therefore, the HALO implementations consistently achieve the maximum performance portability score of 1.0 across all the kernels and devices. Due to the large performance penalty, the hardware-agnostic OpenCL implementations suffer from a low performance portability score ranging from 1.2e-6 to 0.62, indicating unstable and poor performance portability. To justify a practical solution to hardware-agnostic programming for future heterogeneous HPC systems, an average performance portability score of at least 0.95, indicating true performance portability, is needed.

Table \ref{table: sw_overhead} shows the software overhead ratio of HALO 1.0. 
It is shown that the HALO overhead ratio (T1/T4) ranges from 0.0001\%-0.0044\%, 0.0004\%-0.0038\%, and 0\%-0.0015\% for the CPU, GPU, and FPGA, respectively, with the smallest overhead belonging to the matrix-matrix multiplication kernel for all platforms. This is due to its high computational complexity of \(O(n^3)\). The negligible overhead of HALO 1.0 is the key to assuring that the separation between hardware-agnostic host programming and hardware-specific kernel optimization is a viable approach to enabling true performance portability of a host application while maximizing the reusability of kernel codes, which takes the critical first step in minimizing the hardware dependency of HPC applications.

\section{Conclusion and Future Work}
We envision that future HPC will be extreme-scale heterogeneous HPC that integrates various domain- and application-specific accelerators at a massive scale. Application development methodology must shift from a monolithic hardware-aware approach to a performance-portable hardware-agnostic one. Our experiment results show that OpenCL-based solutions suffer from unstably poor performance portability, which is a deal-breaker to heterogeneous HPC. Differently, HALO 1.0 allows the same hardware-agnostic host application, without any change, to run across all the computing devices with negligible performance degradation compared to hardware-optimized implementations, providing a viable solution to hardware-agnostic programming with true performance portability. The unified host code control flow and accelerator interfaces with the novel parent-child ranking system defined by C$^2$MPI are the keys to enabling hardware agnosticism with transparent interoperability and establishing a clear boundary between the responsibility of DMEs and HMEs as well as between hardware-agnostic and hardware-specific software development. The open-ended, extensible software architecture of HALO 1.0 is the key to allowing the inclusiveness required for ensuring prolonged performance portability following the rapid advancement of accelerator technologies. Future work will address the scalability and interoperability perspectives of the proposed solution based on multiple host nodes.



\bibliographystyle{elsarticle-num}
\bibliography{references.bib}





\end{document}